\documentclass[aps,pre,amsmath,amsfonts,floatfix,superscriptaddress,nofootinbib,twocolumn]{revtex4-1}
\pdfoutput=1
\usepackage{mathrsfs} \usepackage{amssymb} \usepackage{graphicx,color} \usepackage{bbm} \usepackage{multirow}
\usepackage{bm} \usepackage{mathrsfs} \usepackage{commath} \usepackage{stmaryrd} \usepackage{color} \usepackage{stmaryrd}
\usepackage{tikz}\usetikzlibrary{shapes,arrows,positioning,calc}\usepackage{comment}
\usepackage{hyperref}  % for useful hyperlinks
\hypersetup{
    colorlinks,
    linkcolor={blue!70!black},
    citecolor={green!70!black},
    urlcolor={red!70!black}
}

\let\a=\alpha \let\b=\beta  \let\d=\delta
   
 \let\m=\mu   
\let\s=\sigma   
   \let\G=\Gamma
\let\D=\Delta   
   
 \let\r=\rho  \let\io=\infty

 \def\VV{{\cal V}}

\def\dd{\mathrm{d}}

\def\to{\rightarrow} \def\la{\left\langle} \def\ra{\right\rangle}

\newcommand{\beq}{\begin{equation}} \newcommand{\eeq}{\end{equation}}
 \newcommand{\wt}{\widetilde}

\def\rs{r^*} \def\rt{\tilde{r}} 

\begin{document}

\title{Approximate scale invariance in particle systems: a large-dimensional justification}

\author{Thibaud Maimbourg}\email{thibaud.maimbourg@lpt.ens.fr}
\affiliation{Laboratoire de Physique Th{\'e}orique, {\'E}cole Normale Sup{\'e}rieure - PSL Research University, CNRS, Universit{\'e} Pierre et Marie Curie - 
   Sorbonne Universit{\'e}s, 24 rue Lhomond, 75005 Paris, France}

\author{Jorge Kurchan}\email{kurchan.jorge@lps.ens.fr}
\affiliation{Laboratoire de Physique Statistique, {\'E}cole Normale Sup{\'e}rieure - PSL Research University, CNRS, Universit{\'e} Pierre et Marie Curie - 
   Sorbonne Universit{\'e}s, Universit{\'e} Paris Diderot - Sorbonne Paris Cit{\'e}, 24 rue Lhomond, 75005 Paris, France}

\begin{abstract}
Systems of particles interacting via inverse-power law potentials have  an invariance with respect to changes
in length and temperature, implying a correspondence 
in the dynamics and thermodynamics between  different `isomorphic'  sets of temperatures and densities. 
In a recent series of works, it has been argued that such correspondences hold to a surprisingly good approximation
in a much more general class of potentials, an observation that summarizes many properties that have been observed in the past.
In this paper we show that such relations are exact in high-dimensional liquids and glasses,
a limit in which the conditions for these mappings to hold become transparent. The special role 
played by the exponential potential is also confirmed.
\end{abstract}

\maketitle

We consider a simple liquid in $d$-dimensional space at temperature $T$ and number density $\rho$.
Two state points $(T_1,\rho_1)$ and $(T_2,\rho_2)$ are \textit{isomorphic} if there is a rescaling of the coordinates of all particles
 $ {\bf r}_i^{(2)} = (\rho_1/\rho_2)^{1/d} {\bf r}_i^{(1)}$  that makes their Boltzmann factors proportional
\begin{equation}
 e^{-U({\bf r}_1^{(1)}, \cdots , {\bf r}_N^{(1)})/T_1}=C_{12}\, e^{-U({\bf r}_1^{(2)}, \cdots , {\bf r}_N^{(2)})/T_2}
\end{equation}
where $C_{12}$ depends only upon the state points 1 and 2, not on the microscopic configurations. This implies a certain number of correspondences \textit{both} in the thermodynamic and dynamic quantities 
between the two states.  Inverse-power law potentials $V_{\rm IPL}(r)\propto r^{-n}$ are the only potentials for which the isomorphic property exactly
 holds, for all points lying on curves given by $\rho^{n/d}/T=$ constant, called \textit{isomorphs}. It is instead only an approximation for all the other potentials, and  yet, surprisingly enough,
 it turns out \cite{BSD14,gnandyreI,gnandyreII,gnandyreIII,gnandyreIV,gnandyreV} that there is a wide class of situations in  which it is a very good one.
This observation led to a number of {\em a posteriori} explanations and interpretations of facts that had been previously observed~\cite{BSD14,gnandyreIV,gnandyreV}, such as striking similarities in structure and dynamics of some model 
liquids~\cite{GRMN03,SHNL11,FL13}, phenomenological rules along freezing or melting lines~\cite{U65,KCM11,A34,HV69,B83}, or observations that a single static quantity controls dynamic properties~\cite{R77,PET11,YA03,Go09}.
It can also be used to rule out theories that are incompatible with these invariances~\cite{gnandyreIV}, or make novel predictions~\cite{XTCDN14,LBD14}.

In this paper we show that the existence of isomorphs in the $(T,\rho)$ plane  actually becomes strictly true in the large-dimensional limit for
a wide and well-defined class of potentials. This holds for the liquid as well as the glassy regions of the phase diagram.
We proceed in three steps: first we show this in detail for the equilibrium properties of the liquid phase, then we show how the same manipulations
allow one to prove it for the  dynamics, and finally we outline a derivation for the equilibrium (`landscape') properties within
the glass phase. The main argument is the following: `soft' interactions in large dimensions are such that particles that are too close are exponentially few in number,
while those that are too far interact exponentially weakly: all in all there is a typical distance (to within $1/d$) where interactions are relevant.\\
 
\paragraph*{\bf High-dimensional liquids --}

We  wish to focus on the large $d$ behaviour of a system of $N$ particles interacting via a pair potential $V(|{\bf r}|)$.
% We will use the setting of~\cite{MKZ16,KMZ16}, confining the particles in a `box' consisting in the surface of a 
% $d+1$-dimensional sphere of radius $R$, very large compared to the interparticle distance; the usual $d$-dimensional Euclidean space is recovered when 
% $R\rightarrow\infty$. 
In order to have a well-defined large-dimensional limit, we shall assume that the potential scales as 
\begin{equation}\label{eq:pot}
V(r)=e^{-d A(r)} 
\end{equation}
%\textcolor{red}{(remove this sentence?) In this paper we shall restrict ourselves to a monotonically decreasing form of $V(r)$, implying an increasing function $A(r)$ (this is not strictly necessary)}.
The precise conditions on  $V(r)$ will be derived below.
 
\begin{figure}[h]
\begin{center}
\includegraphics[width=7cm]{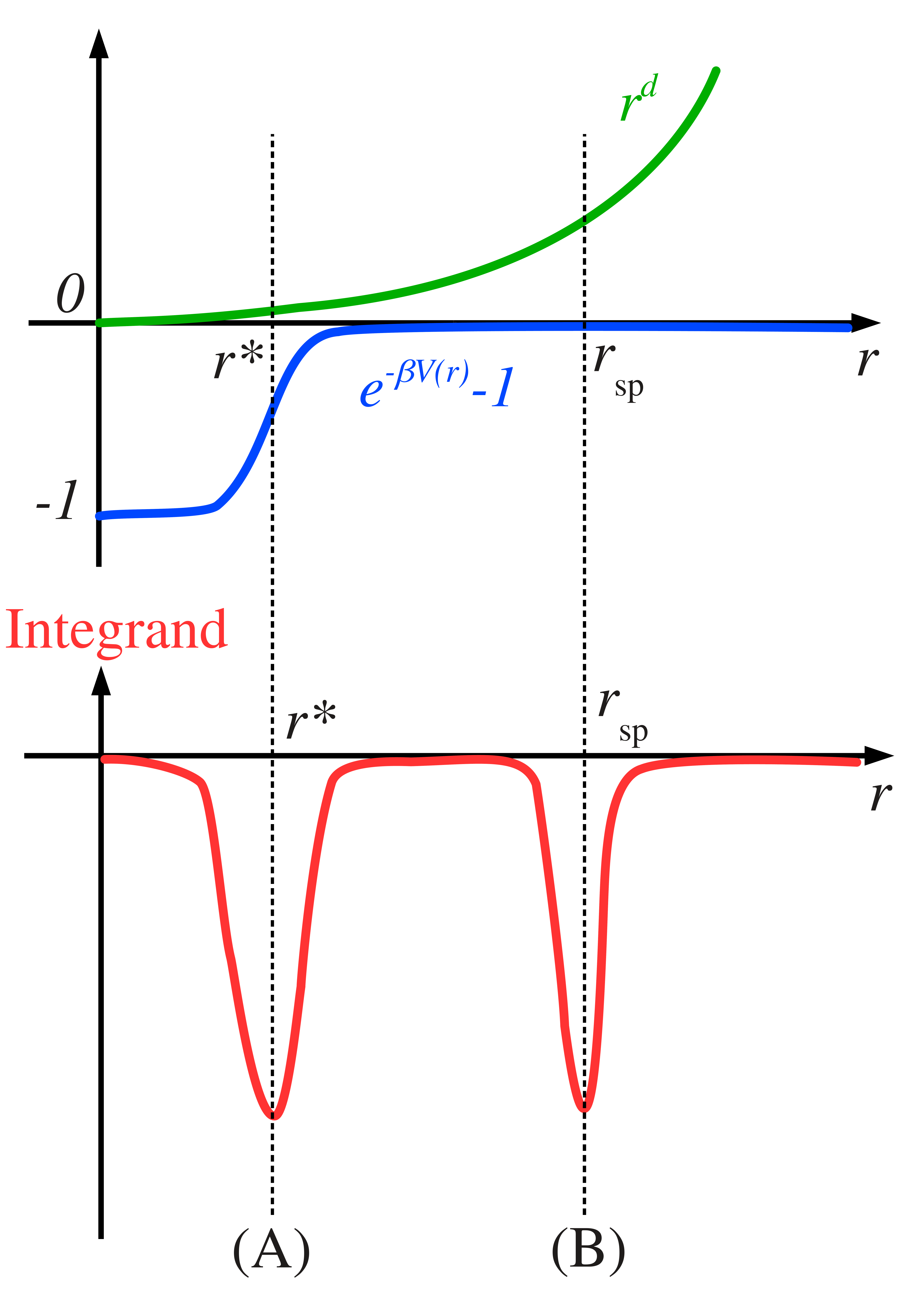}
\end{center}
\caption{The two dominant regions in the integrand of the Mayer integral for large $d$. The blue curve is the Mayer function, the green one is the power law $r^d$ arising from the measure, and the red one is the 
product of the two giving the integrand.}
\label{fig:vir}
\end{figure}

For large $d$ only the first two terms of the virial expansion contribute; the free energy and the pressure of the liquid are respectively given by~\cite{WRF87,FP99,hansen,MH61,KMZ16}
\begin{equation}\label{eq:FP}
\begin{split}
  \frac{\beta F}{N}&=\frac{\beta (F_{\rm IG}+F_{\rm ex})}{N}=\ln \rho -1 -\frac{\rho}{2}\int \mathrm{d} \mathbf{r}\, \left(e^{-\beta V(r)}-1\right)\\
   \frac{\beta P}{\rho}&=1 -\frac{\rho}{2}\int \mathrm{d} \mathbf{r}\, \left(e^{-\beta V(r)}-1\right)
\end{split}
\end{equation}
where $\rho=N/V$ is the particle density. The free energy has an ideal gas contribution $F_{\rm IG}=N(\ln\rho-1)/\beta$ and an excess contribution $F_{\rm ex}$ given by the last
term. Let us focus on the Mayer integral in the excess part:
\begin{equation}
  \int \mathrm{d} \mathbf{r}\, \left(e^{-\beta V(r)}-1\right)=\Omega_d \int_0^\infty \mathrm{d} r\,r^{d-1} \left(e^{-\beta V(r)}-1\right)
\end{equation}
where $\Omega_d$ is the area of the unit sphere. The integrand looks as in fig.~\ref{fig:vir}.\\

Set an effective radius $r^*$ such that
\begin{equation}\label{eq:tildebeta}
\beta V(r^*) \equiv \widetilde\beta
\end{equation}
with $\widetilde\beta$ any number\footnote{If $\widetilde\beta=Be^{bd}$ and $V(r)=V_0e^{-dA(r)}$ with $B$ and $V_0$ subdominant, then $r^*$ is given by 
$A(r^*)=b$ and $\widetilde\beta=BV_0$. The case $\beta=O(1)$ was the case studied in~\cite{MKZ16,KMZ16} which is easily extended to other scalings of temperature as analyzed here.} of order $O(1)$. 
This defines $r^*$ up to $O(1/d)$, independently of the value of 
$\widetilde\beta$ as long as it remains of order one. 
% In the case $\beta=O(1)$, $r^*$ is precisely defined by the condition $A(r^*)=0$. 
% %The precise value of $\widetilde\beta$ matters and is just an
% %adimensionalization of the temperature. 
% This was the case studied in~\cite{MKZ16,KMZ16}.}).
We distinguish three regions for the Mayer function $f(r)=e^{-\beta V(r)}-1$:
\begin{itemize}
 \item {$r<r^*$}, so that $\underset{d\rightarrow\infty}{\lim}f(r)=-1$. Then essentially the integrand is $\sim -r^d$, and in large dimensions only the right boundary $r \rightarrow r^*$
 dominates.
\item {\bf $r \sim r^*$} up to $O(1/d)$, where $0<f(r)<-1$.  With our scaling of the potential in the large $d$ limit, this region has a $O(1/d)$ extension and,
setting $r=r^{*}(1+\tilde{r}/d)$, one may compute the contribution of this region (A), as denoted in fig.~\ref{fig:vir}, as
\begin{equation}\label{eq:fcalc}
\begin{split}
   &\Omega_d \int_{\rm (A)} \mathrm{d} r\,r^{d-1} \left(e^{-\beta V(r)}-1\right)= \\
   &\mathcal{V}_d(r^{*})\int_{-O(d)}^{O(d)} \mathrm{d} \tilde{r}\, \left(1+\frac{\tilde{r}}{d}\right)^{d-1} \left[e^{-\beta V(r^{*}+\frac{r^{*}}{d}\tilde{r})}-1\right]\sim\\
   &\mathcal{V}_d(r^{*})\int_{-\infty}^\infty \mathrm{d} \tilde{r}\, e^{\tilde{r}}\left[e^{-\widetilde\beta \exp(-r^*A'(r^*)\tilde{r})}-1\right]
\end{split}
\end{equation}
with $\widetilde\beta$ defined in~\eqref{eq:tildebeta} and  $\mathcal{V}_d(\sigma)=\Omega_d \sigma^d/d$ is the $d$-dimensional volume of the sphere of diameter $ \sigma$.
We have expanded $A(r)$ around $r^{*}$ and kept non-vanishing orders when $d$ goes to infinity.  
This development only makes sense if %the exponential is decreasing, \textit{i.e.} if
\begin{equation}\label{eq:condrs}
 A'(r^{*})>\frac{1}{r^{*}} \; ,
\end{equation}
which ensures the convergence of the integral, and in this case only a range of values around $r^*$ of width $O(1/d)$ contributes, a crucial fact.
\item {\bf $r>r^*$}. These  may give  a further contribution. Here, we have  $\underset{d\to\infty}{\lim}f(r)=0$
and  one can expand the Mayer function as $f(r)\sim- \beta V(r)$. A saddle-point
evaluation of the Mayer integral in this region is possible, obtained
by maximizing the exponent $\ln r-A(r)$ in a point  $r_\mathrm{sp}$  
given by  (see region (B) in fig.~\ref{fig:vir}):
\begin{equation}\label{eq:condrsp}
 A'(r_\mathrm{sp})=\frac{1}{r_\mathrm{sp}}
\end{equation}
As usual, the fluctuations around this saddle-point  are of order $1/\sqrt d$.
\end{itemize}

\paragraph*{\bf A second order pseudo-transition --}

At each density and temperature, it may be the case that for large $d$ the integral is dominated
either by the region around $r^*$, the region around a saddle point $r_\mathrm{sp}$ or by $r \rightarrow \infty$. The latter case
corresponds to long-range potentials, and we shall not consider it here.
If condition (\ref{eq:condrs}) is satisfied, the contribution around $r^*$ is well defined. The question remains if
there exists some saddle-point value $r_\mathrm{sp}$, \textit{i.e.} if~\eqref{eq:condrsp} defines an absolute maximum of the potential for $r>r^*$, that dominates.
Assume that there is such a  value, and let us compare the contributions
 \begin{equation}
 \begin{split}
  &(r^{*})^d(e^{-\widetilde\beta   \exp(-r^*A'(r^*)\tilde{r})    }-1)\hskip15pt\textrm{around}~r^*\ ,\\
  &r_\mathrm{sp}^d\left[\exp\left(-\widetilde\beta e^{d(A(r^{*})-A(r_\mathrm{sp}))}\right) -1\right]\sim\\
  &-\widetilde\beta r_\mathrm{sp}^d e^{d(A(r^{*})-A(r_\mathrm{sp}))}\hskip15pt\textrm{around}~r_\mathrm{sp} \ .
 \end{split}
 \end{equation}
We have, for large $d$, that the neighbourhood of $r^*$ dominates if
\begin{equation}
\left[ \ln r -  A(r)\right]^{r^*}_{r} >0 \hskip15pt \textrm{for}\;r>r^*
\label{condition2}
\end{equation}
which holds \textit{e.g.} if $\ln r - A(r)$ is a monotonically decreasing function for $r\geqslant r^*$. 

Some potentials, such as  the Gaussian one $V(r)=e^{-ad r^2/2}$ have a low-temperature  `rather hard'  regime
where $r^*$ dominates and a high temperature `very soft'  regime  where $r_{sp}$  dominates.
This is a second order {\em pseudo}transition, it exists only in the limit $d \rightarrow \infty$.
The very soft `phase' has constant energy, as can be easily seen by differentiating (\ref{eq:FP}), and its interpretation is clear: the high-temperature expansion
is exact to first order in $\beta$, each particle strongly interacts with many others, the exponential number of those that are at a distance $r_{sp}$ dominate the interactions.
The transition temperature between these two `phases' is precisely the value of temperature where $\left[ \ln r -  A(r)\right]^{r^*}_{r_{sp}} =0$.
Isomorphs exist for the  `rather hard potentials' `phase', we shall not consider the case 
 in which a typical distance  $r_\mathrm{sp}> r^*$ dominates (`very soft spheres') -- which has never been studied in the high-dimensional limit, 
but may be easily treated, dynamically and statically, to first order in the large-temperature expansion.\\

% 
%\begin{figure}[h!]\label{fig:rsrsp}
% \begin{tabular}{cc}
%%  \includegraphics[width=6cm]{lnA1.jpg} & \includegraphics[width=6cm]{lnA2.jpg}\\
%  (a) & (b)
% \end{tabular}
% \caption{Cases where both peaks (A) and (B) could contribute to the Mayer integral: the one that prevails is (a) $r^{*}$ (b) $r^{*}p$.}
%\end{figure}
 
\paragraph*{\bf Isomorphs and effective potential --}\label{sec:isomorph}

In the `rather hard' regimes where $r^{*}$ dominates the Mayer integral, from~\eqref{eq:fcalc} the free energy reads:
\begin{equation}\label{eq:Feff}
  \frac{\beta F}{N}=\ln \rho -1 -\frac{\widetilde{\varphi}}{2}\int_{-\infty}^\infty \mathrm{d} \tilde{r}\, e^{\tilde{r}}\left[e^{-\widetilde\beta e^{-\tilde{r}/\alpha}}-1\right]
\end{equation}
where 
\begin{equation}\label{eq:phialpha}
 \widetilde{\varphi}\equiv\rho\mathcal{V}_d(r^*)\,,\hskip15pt 1/\alpha\equiv r^{*} A'(r^{*})>1
\end{equation}
The term $e^{\tilde{r}}$, related to the linear part of the effective potential in~\cite{KMZ16,MKZ16}, is an entropic `driving force',
$\widetilde \beta$ is an effective inverse temperature and $\widetilde\varphi$ an effective packing fraction.
 We have discovered naturally that the effective potential is a pure decaying exponential $V_{\rm eff}(r)=V(r^*)\exp(-r^*A'(r^*)\tilde{r})=V(r^*)\exp(-\tilde r/\alpha)$
similar to the `building blocks' of Dyre, Schr{\o}der et al~\cite{BSD14,BD14}. The value $\alpha$ gives its `hardness' and 
depends upon the value of $r^{*}$ and  the original potential through $A'$.\\

We now show under which conditions two systems at different state points are related through a scaling transformation. If one shifts in the virial
term of equation~\eqref{eq:Feff} $\tilde{r}\to\tilde{r}+c$, which corresponds to an order $1/d$ shift of $r^{*}$, one gets
\begin{equation}\label{eq:shiftc}
\begin{split}
 &\frac{\widetilde{\varphi}}{2}\int_{-\infty}^\infty \mathrm{d} \tilde{r}\, e^{\tilde{r}}\left[e^{-\widetilde\beta e^{-\tilde{r}/\alpha}}-1\right]=\\
 & \frac{\widetilde{\varphi}e^c}{2}\int_{-\infty}^\infty \mathrm{d} \tilde{r}\, e^{\tilde{r}}\left[e^{-\widetilde\beta e^{-c/\alpha}e^{-\tilde{r}/\alpha}}-1\right]
\end{split}
 \end{equation}
Hence, a sufficient condition for two systems at different state points 1 and 2 to have the same excess free energy is that there exists $c$ such that
\begin{equation}\label{eq:shift}
\left\{ 
\begin{split}
  &r^{*}_1A'(r^{*}_1)= r^{*}_2A'(r^{*}_2)=1/\alpha\\
  &\widetilde\varphi_1e^c=\widetilde\varphi_2=\widetilde\varphi\\
  &\widetilde{\beta}_1e^{-c/\alpha}=\widetilde{\beta}_2=\widetilde\beta
\end{split}
\right.
% \hskip15pt
%\Leftrightarrow \hskip15pt\left\{
%\begin{split}
%  r^{*}_1A_1'(r^{*}_1)&= r^{*}_2A_2'(r^{*}_2)=1/\a\\
%  \r_1\mathcal{V}_d(r^{*}_1)e^c&=\r_2\mathcal{V}_d(r^{*}_2)=\widetilde\varphi\\
%  \b_1 V_1(r^{*}_1)e^{-c/\a}&=\b_2 V_2(r^{*}_2)=\widetilde\b
%\end{split}
%\right. 
\end{equation}
These  conditions mean that one may set
$r^{*}_1=r^{*}_2=r^{*}$ (which determines $1/\alpha=r^{*} A'(r^{*})$) and that they may be mapped by
\begin{equation}
\label{cos}
\left\{ 
\begin{split}
  &\rho_1e^c=\rho_2\\
  &\beta_1 e^{-c/\alpha}=\beta_2 
\end{split}
\right. \hskip15pt
\Leftrightarrow\hskip15pt 
\frac{\rho_1^{\frac1\alpha}}{T_1}=\frac{\rho_2^{\frac1\alpha}}{T_2}
\end{equation}
via {\em  a value of $c$ of  sub-exponential order in $d$} which parametrizes the line
\begin{equation}\label{eq:iso}
 \frac{\rho^{\frac1\alpha}}{T}=\rm constant
\end{equation}
Along these lines, the partition function between two points scales changes by a purely geometric, model-independent factor:
\begin{equation}
\beta_2F_2-\beta_1F_1 =N \ln \frac{\rho_2}{ \rho_1}
\end{equation}
as may be easily verified using the transformations (\ref{cos}) in (\ref{eq:FP}). This factor is only due to ideal gas contributions; the excess free energy, which contains
all potential-dependent static properties, is invariant under this transformation, explaining a number of static discoveries described in~\cite{gnandyreIV}.

The correspondence  is then valid in any phase diagram window where density and temperature are rescaled even by large factors,
provided they are smaller than  exponential in $d$ (\textit{i.e.} that temperatures are not rescaled exponentially in $d$), so that
all points  have the same $r^*$. 
% Note that all reference to the (arbitrary) number $\widetilde \beta$ has disappeared -- it only sets the constant that labels each line, but
% the lines are the same whatever the choice.
Greater (exponential) changes of  parameters modify the value of $r^*$, and through it the hardness of the potential $1/\alpha=A'(r^*)r^*$. This is the case even if one follows the same curve given 
by~\eqref{eq:iso}, for instance by performing an exponential compression and heating of the system. The mappings are in this case only approximate, except for the case where $1/\alpha=A'(r^*)r^*=$ constant $\forall\,r^*$, \textit{i.e.} the inverse-power law
potential. 
%Note that, however, care must be taken when considering  too high densities since the convergence of the virial series is not guaranteed above 
%$\widetilde \varphi\simeq (e/2)^{d/2}$~\cite{FP99}, and the results presented here may therefore not hold.}

Note that, generalizing the above relations, one can map two systems at different state points 1 and 2 \textit{and with different interaction potentials} given by their exponent 
$A_1$ and $A_2$. In this case only the first condition in~\eqref{eq:shift} is modified and reads $r^{*}_1A_1'(r^{*}_1)= r^{*}_2A_2'(r^{*}_2)=1/\alpha$. The 
same hardness $1/\alpha$ will generally imply different effective radii $r^*_1\neq r^*_2$, so that the two other equations become
$\widetilde\varphi_1e^c=\widetilde\varphi_2=\widetilde\varphi =  \rho_1 \mathcal{V}_d(r_1^*)e^c=\rho_2 \mathcal{V}_d(r_2^*)$ and
$\widetilde{\beta}_1e^{-c/\alpha}=\widetilde{\beta}_2=\widetilde\beta$. The invariant curves will thus have the more general equation 
$\widetilde\beta[{\rho\mathcal{V}_d(r^*)}]^{1/\alpha}=$ constant. \\
 
% {\color{red}
% \paragraph*{The special role of exponential potentials}\label{sec:mother}
% 
% In accordance with Ref \cite{BD14}, we find that exponential potentials indeed play a special role.
% We may write:
% \begin{equation}
%  V(r)=e^{-d A(r)}=\int_0^\infty\mathrm{d} a\,e^{-dar}e^{dg(a)}\sim e^{d[g(a^*)-a^* r]}
% \end{equation}
% or, for a finite sum, 
% \begin{equation}
%  V(r)=\sum_i e^{-d a_i r}e^{d g_i}\sim e^{d\,\underset{i}{\max}(g_i-a_i r)}
% \end{equation}
%  again as a Legendre transform. For every value of external parameters, there is a single $r^*$, and hence a  single $a^*$  that dominates.
%  Then, only one exponential dominates, with prefactor $r^* A'(r^*)$. \\
% \textbf{Example of the Lennard-Jones potential}
% 
% }

\paragraph*{\bf Dynamics --}
 The  dynamical equations, exact in the limit of high dimensions, have been recently derived~\cite{MKZ16,KMZ16}. 
 It turns out that, as is usual in all large $d$ computations, the system breaks down into the dynamics of a single degree of freedom
 $\tilde r(t)$ with colored  noise $\xi$ and friction  $\int_{t_0}^t \mathrm{d} s\, \beta^2 M(t-s) \dot{\tilde{r}}(s)$, defined by a memory kernel  $\beta^2M(t)$ that have to be derived self-consistently. 

We define the potential centered around $r^*$ as $\bar V(\tilde r)=V[r^*(1+\tilde r/d)] $. Similarly to the statics, the average relative distance between nearest neighbours 
turns out to be $r(t)=r^*(1+\tilde r(t)/d)$. For equilibrium dynamics, where the initial condition at $t_0$ is picked with the canonical equilibrium probability, 
the self-consistent equations for finite times $t$ may be written as~\cite{MKZ16,KMZ16}:
\begin{equation}\label{eq:dynpro}
\begin{split}
&\beta\widehat m   \ddot{\tilde{r}}(t) +\beta\widehat \gamma \dot{\tilde{r}}(t)
= 1- \int_{t_0}^t \mathrm{d} s\, \beta^2 M(t-s) \dot{\tilde{r}}(s)\\
&\hskip80pt- \beta \bar V'(\tilde{r}(t))  + \xi(t)\,, \\
&\langle \xi(t)\rangle=0\,,~\langle \xi(t) \xi(t') \rangle =2\beta\widehat\gamma\delta(t-t')+\beta^2 M(t-t')\,, \\
& \beta^2M(t-t')=\frac{\widetilde\varphi}{2d}\int\mathrm{d} \tilde{r}_0\,e^{\tilde{r}_0-\beta \bar V(\tilde{r}_0)}\langle \beta \bar V'(\tilde{r}(t))\beta \bar V'(\tilde{r}(t'))\rangle
\end{split}
% \begin{split}
% \beta\widehat m   \ddot{\tilde{r}}(t) +\beta\widehat \gamma \dot{\tilde{r}}(t)
% =& -\frac{\tilde r(t)}{\Delta_{\rm liq}}- \int_{t_0}^t \mathrm{d} s\, \beta^2 M(t-s) \dot{\tilde{r}}(s)\\
% &- \beta \bar V'(\tilde{r}(t)+ \lambda)  + \xi(t) \\
% \textrm{with}\hskip15pt\langle \xi(t)\rangle=&0\,,\hskip15pt \langle \xi(t) \xi(t') \rangle =\beta^2 M(t-t')\,, \\
% \beta^2M(t-t')  = \frac{ \widetilde\varphi}{2d} &\int \mathrm{d}\lambda\, e^{\lambda-\Delta_{\rm liq}/2} \\
% &\times\langle \beta \bar V'(\tilde{r}(t)+\lambda) \beta \bar V'(\tilde{r}(t')+\lambda) \rangle
% \end{split}
\end{equation}
where $\widehat m=(r^{*})^2m/2d^2$ and $\widehat \gamma=(r^{*})^2\gamma/2d^2$, $m$ being the physical mass and $\gamma$ the coupling to the bath. If we set $\gamma=0$ we have the purely Newtonian case, and with $m=0$ the overdamped Brownian case.
Out-of-equilibrium dynamics may also be considered (see next section).

For the purely Newtonian case, we adimensionalize these equations by setting $\tilde t=t/\sqrt{\widehat m\beta}$.
We expand as before the potentials around $r^*$ as $\beta V\left(r^{*}[1+\tilde{r}(t)/d]\right)=\widetilde\beta e^{-\tilde{r}(t)/\alpha}$, and get:
\begin{equation}
\begin{split}
\frac{\mathrm{d}^2\tilde{r}}{\mathrm{d}\tilde{t}^2}=& - \int_{\tilde t_0}^{\tilde t} \mathrm{d}\tilde s\, \widetilde M(\tilde t-\tilde s) \frac{\mathrm{d}\tilde{r}}{\mathrm{d}\tilde{s}}(\tilde s)
+\frac{\widetilde\beta}{\alpha} e^{-\tilde{r}(\tilde{t})/\alpha}  + \widetilde\xi(\tilde t) \\
\textrm{with}&\hskip15pt\langle \widetilde\xi(\tilde t)\rangle=0\,,\hskip15pt \langle \widetilde\xi(\tilde t)\widetilde \xi(\tilde t') \rangle =\widetilde M(\tilde t-\tilde{t}')\,, \\
\widetilde M(\tilde t-\tilde t')  =& \frac{ \widetilde\varphi}{2d}\int\mathrm{d} \tilde{r}_0\,e^{\tilde{r}_0-\beta \bar V(\tilde{r}_0)}
\left\langle \frac{\widetilde\beta}{\alpha} e^{-\tilde{r}(\tilde t)/\alpha}\frac{\widetilde\beta}{\alpha} e^{-\tilde{r}(\tilde{t}')/\alpha}\right\rangle
\end{split}
\end{equation}
The tilde variables are rescalings in the new time units, \textit{e.g.} $\widetilde M(\tilde t)=\beta^2 M(\tilde t\sqrt{\widehat m\beta})$ (nevertheless we kept the same symbol for $\tilde r$ to simplify
the notation). Now, performing the shift $\tilde{r}(t)\to\tilde{r}(t)+c,\,\forall t$, we find exactly the same rescalings of parameters as in the static liquid phase computations, and once again
the isomorphs are given by~\eqref{eq:iso}.

 For the purely Brownian case,  we instead adimensionalize these equations by setting $\tilde t=t/\beta\widehat \gamma$, 
rescale the  variables  in the new time units, \textit{e.g.} $\widetilde M(\tilde t)=\beta^2 M(\tilde t\beta\widehat\gamma)$, and check that the translation $\tilde{r}(t)\to\tilde{r}(t)+c$  has the same effect as before.
Interestingly, for the mixed case with friction and inertia, going from one state point to the other changes the ratio of inertial to bath intensities through an additional parameter $\widehat\gamma\sqrt{\beta/\widehat m}$. 
If this parameter is large (respectively, small) when $d\to\infty$ then the equation reduces to the Brownian (respectively, Newtonian) case. The mixed dynamics is not fully invariant, although the correspondence 
is simple, and this only affects high-frequency properties. \\

% Note that the unit of energy is $1/\b$, the unit of length is given by $r^{*}$ and the unit of time is $r^{*}\sqrt{\beta m}$ for Newtonian dynamics and $\beta\gamma(r^{*})^2$ for Brownian dynamics, 
% as in~\cite{gnandyreIV}. The unit of length is given there by $\r^{-1/3}$. This is actually the same as an effective proper space per particle. If we impose an effective packing fraction $\widetilde\varphi=\r\mathcal{V}_d(r^{*})$ of order 1
% we must have $r^{*}\sim(\r\Omega_d/d)^{-1/d}$ to leading order; if we ignore the geometrical factor due to the sphericity of the spheres, which is a numerical constant in finite dimension, we have that 
% $r^{*}\propto\r^{-1/d}$.\\
% All of this is not surprising because it is strictly dictated by dimensional analysis since there are very few relevant parameters.
%Here the situation allows to make a closer contact to the static liquid phase computations, since the invariance amounts to shifting $\tilde{r}(t)\to\tilde{r}(t)+c,\,\forall t$.\\

\paragraph*{\bf Glassy phases --}

Glassy phases of particle systems in the regime considered here have been studied in~\cite{PZ10,KPZ12,KPUZ13,CKPUZ14,nature,KMZ16,RUYZ14}. We will use the setting of~\cite{MKZ16,KMZ16}, confining the particles in a `box' consisting in the surface of a 
$d+1$-dimensional sphere of radius $R$, very large compared to the interparticle distance; the usual $d$-dimensional Euclidean space is recovered when 
$R\rightarrow\infty$. Replacing the effective hard sphere radius $\sigma$ in~\cite{KMZ16} by $r^{*}$ here, the replicated free energy is given in terms of a replica matrix 
$Q_{ab}= \frac{2d}{(r^*)^2}{\bf r}_a \cdot {\bf r}_b$ that encodes the distance between the $n$ replicas of a particle $\mathbf r$:
\begin{equation}
\label{eq:SSfinal}
\begin{split}
&\frac{\beta F(\hat Q)}{N} =n(\ln N -1) -\frac{d}2 n \ln\left( \frac{\pi e (r^{*})^2}{d^2}\right) - \frac{d}2  \ln\det\hat Q \\
&-  \frac{\widetilde\varphi }{2} \int\mathrm{d}\lambda \mathcal{D}_{\hat Q}\bar{\tilde{r}} \, e^{\lambda-\Delta_{\rm liq}/2} \, \left[e^{-\beta\sum_{a=1}^n V\left(r^{*}\left(1+ \frac{\tilde{r}_a + \lambda}{d} \right)\right)}- 1 \right]
\end{split}
\end{equation}
% \end{widetext}
% \begin{floatequation}
% \mbox{\textit{see eq.~\eqref{eq:SSfinal}}}
% \end{floatequation}
where 
\begin{equation}
 \mathcal{D}_{\hat Q}\bar{\tilde{r}} \equiv \frac{e^{-\frac12\tilde{r}_a {Q}^{-1}_{ab} \tilde{r}_b}}{(2\pi)^{n/2}\sqrt{\det\hat Q}}\prod_{a=1}^n \mathrm{d}\tilde{r}_a 
 \end{equation}
$\Delta_{\rm liq}\equiv 2d(R/r^*)^2$ represents the size of the `box'. 
Even before making any ansatz for $\hat Q$, we may show that a mapping exists by defining $\widetilde\beta$ as in~\eqref{eq:tildebeta} and $\widetilde\varphi$ as in~\eqref{eq:phialpha}.
Expanding  each one of the Mayer terms around $r^*$ as before, we obtain 
the same equations as in the static case studied in refs.~\cite{KMZ16,KPZ12,KPUZ13,CKPUZ14,nature}, but with the exponential potential $ e^{-(\tilde{r}_a + \lambda)/\alpha}$. 
Using the translation $\lambda\rightarrow\lambda+c$, we conclude that  condition~\eqref{eq:iso} defines isomorphs in the glassy region of the phase diagram of the system as well. 
This in itself implies that the entire structure of metastable states is the same along points in an isomorph. 

The out-of-equilibrium dynamics can be derived along the same lines as~\cite{MKZ16}. 
% , giving similar self-consistent equations governed by the one-dimensional off-equilibrium
% dynamics of $\tilde r(t)$. 
The dynamical action bears a formal analogy with the replicated free energy~\eqref{eq:SSfinal} where, respectively, time and replica index play a 
similar role~\cite{KMZ16}, and once again a shift of $\lambda$ gives the equation of the same invariant curves~\eqref{eq:iso}.\\

\paragraph*{\bf Virial-energy correlations --}

In~\cite{BSD14,gnandyreIV} a simple measure of the goodness of scaling relations was introduced as follows. The energy 
$U=\sum_{i<j}V(|\mathbf{r}_i-\mathbf{r}_j|)$ and the so-called virial function \cite{hansen} $W=-\frac 1d \sum_i \mathbf{r}_i\cdot\mathbf{\nabla}_{\mathbf{r}_i}U$  are used to define a
virial-energy correlation coefficient $R\in[-1,1]$ as
\begin{equation}\label{eq:R}
 R=\frac{\langle \Delta W\Delta U\rangle}{\sqrt{\langle (\Delta W)^2\rangle\langle (\Delta U)^2\rangle}}
\end{equation}
where $\Delta W=W-\langle W\rangle$ and brackets denote equilibrium averages. In the case of inverse-power law
 potentials, we have $W\propto U$  and $R=1$. 
 In any practical case, one may numerically compute $R$: a value close to unity is an indication
 of good scaling properties~\cite{gnandyreIV,gnandyreV,BSD14}. 
 
 Here one easily shows that $R=1$ for large dimensions for any potential satisfying the `rather hard' condition.
 This is done by computing directly  $R$ through~\eqref{eq:R}, using equilibrium averages truncated at the lowest order of the virial expansion (see appendix). To leading order:
\begin{equation}\label{eq:Rdio}
 R=-\frac{\int\mathrm{d}\mathbf{r}\,[rV'(r)+d\,V(r)]e^{-\beta V(r)}}{\beta\sqrt{\int\mathrm{d}\mathbf{r}\,r^2V'(r)^2e^{-\beta V(r)}\int\mathrm{d}\mathbf{r'}\,V(r')^2e^{-\beta V(r')}}}
\end{equation}
One can check, by the same reasoning as the one used for the Mayer integral, that these integrals are dominated, under the previous conditions, by the $O(1/d)$ neighborhood of $r^*$ (see appendix). 
Expanding this expression once again around $r^*$ gives $R=1$ identically (see appendix). 
An alternative, quicker way, is to recognize that any potential may be substituted by an inverse-power law potential $V_{\rm IPL}(r)\propto (r^*/r)^{d/\alpha}$ once
the value of $r^*$ is fixed such that $1/\alpha=r^*A'(r^*)$.\\

\paragraph*{\bf Other types of potentials --}

Other potentials can be considered, analyzing the Mayer integral in the same way. The case of a sum of exponential potentials~\eqref{eq:pot} 
with different interaction ranges (measured by $\alpha$) is straightforward: for a given temperature (thus $r^*$), only one of the terms will dominate. The situation is more complex 
if we consider terms with different signs and/or same interaction ranges. The Lennard-Jones (LJ) potential belongs to this class\footnote{A similar discussion can be made with the Weeks-Chandler-Anderson 
potential~\cite{WCA71}.}; its study can readily be done along the same lines and its interest also lies in the fact that it is directly relevant for three-dimensional liquids and 
glasses~\cite{hansen,gnandyreII,gnandyreV,BT09,BT11}. This potential can be generalized in $d$-dimensions as 
\begin{equation}\label{eq:LJ}
 V_{\rm LJ}(r)\propto\left(\frac{r}{\sigma}\right)^{-4d}-\left(\frac{r}{\sigma}\right)^{-2d}
\end{equation}
The Mayer function is very similar: it is $-1$ at short distance, then there is an $O(1)$ part over a range $O(1/d)$ around $\sigma$ where there is a positive bump, and it is exponentially small at 
large distances (see appendix).
\begin{figure}[h!]
\centering
 \includegraphics[width=7cm]{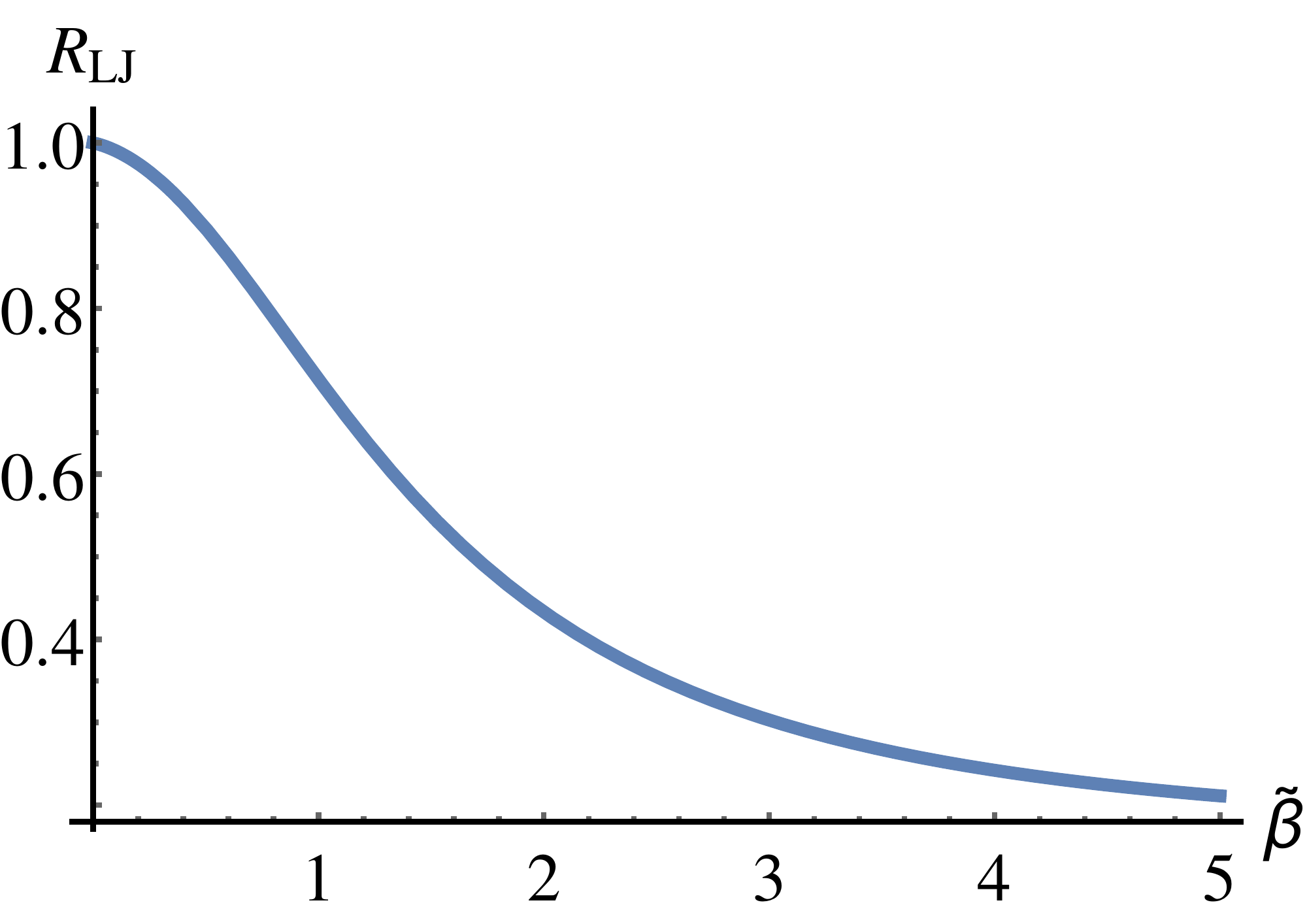}
 \caption{The correlation coefficient $R_{\rm LJ}(\widetilde\beta)$ for the peculiar regime of $O(1)$ temperature.}
 \label{fig:LJ}
\end{figure}
For exponentially high temperatures, one can define a $r^*<\sigma$ where only the repulsive IPL term plays a role; one thus finds trivial isomorphs.
The interesting regime is when the temperature is $O(1)$. Then $r^*$ is defined in the $O(1/d)$ region close to $\sigma$ where attractive and repulsive parts compete. 
We can expand the potential once again setting $r=r^*(1+\tilde r/d)$, giving an effective potential
\begin{equation}
 \beta V_{\rm LJ}(r)\sim\widetilde\beta V_{\rm LJ}^{\rm eff}(\tilde r)=\widetilde\beta(e^{-4\tilde r}-e^{-2\tilde r})
\end{equation}
Here the previous scaling transformations does not provide an invariance since the two exponentials have a different `exponent' $\alpha$. 
One can still demand that the static liquid excess entropy $F_{\rm ex}(\rho,T)=$ constant, defining generally lines. But these 
lines will not have a dynamic (respectively glassy static) counterpart, which would depend upon the times (respectively the replica `blocks') considered. Thus in this regime there are no isomorphs (see footnote \footnote{One 
can write the LJ potential~\eqref{eq:LJ} as~\eqref{eq:pot} but the assumption which is not fulfilled here is that $A(r)$ becomes discontinuous at $r=\sigma$: one cannot expand $A(r)$ around this point and 
the previous analysis does not hold.}).
% One could try to have a dynamic invariance by setting, in reduced units, $M(\tilde t-\tilde t';\beta,\rho)=\textrm{constant}(t,t')$
% (explain it in SUSY also) but the lines $\rho=f(T,\tilde t,\tilde t')$ will depend upon the times considered. Similarly in the statics, the lines where the excess entropy is 
% constant will depend upon the replica indices $(a,b)$ (for non-replica-symmetric order parameter). 
We plot in fig.~\ref{fig:LJ} the virial-energy correlation coefficient in this regime (see appendix), by computing the dominant contribution (around $r^*$) of~\eqref{eq:Rdio}.
% , giving
% \begin{equation}
%  R\underset{d\to\io}{\sim}-\frac{1}{\widetilde\beta}\frac{\int_{-\io}^\io\dd\rt\,(-3e^{-4\rt}+e^{-2\rt}) e^{\rt-\wt\b(e^{-4\rt}-e^{-2\rt})}}
%  {\sqrt{\int_{-\io}^\io\dd\rt\,(-4e^{-4\rt}+2e^{-2\rt})^2 e^{\rt-\wt\b(e^{-4\rt}-e^{-2\rt})}\int_{-\io}^\io\dd\rt\,(e^{-4\rt}-e^{-2\rt})^2 e^{\rt-\wt\b(e^{-4\rt}-e^{-2\rt})}}}
% \end{equation}
% \textcolor{red}{(here a plot of $R$ versus $\widetilde\beta$)}\\
We note accordingly that the $WU$ correlation coefficient is less than one, except in the `infinite' temperature limit $\widetilde\beta\to0$ within this regime. Indeed at high
temperature only the repulsive IPL term is felt by the system. It coincides with trying to shift $r^*$ to lower values where only this term is relevant, and where the 
isomorphs are exact.\\

\paragraph*{\bf Discussion --}

We have shown that whenever the potential $V(r)=e^{-dA(r)}$ satisfies the condition that at the point $r^*$ such that $\beta V(r^*) = O(1)$ one has
$\left[\ln{r}-A(r)\right]^{r^*}_r>0$ for $r>r^*$, then in the large-dimensional limit the parameter space $(T,\rho)$ is foliated with lines (isomorphs), 
where static and dynamic properties coincide once expressed in reduced units. The simple explanation of this fact is that, in large dimensions,
there is a typical interparticle distance that dominates the physics: larger distances have weak interactions, shorter distances are too rare.
The arguments hold for dynamic as well as equilibrium calculations, both in the liquid and in the glass phase. Transition and dynamical
crossover lines follow these isomorphs.

There is numerical evidence that in $d=2$, 3 and 4 in Lennard-Jones systems, both the virial-energy correlation coefficient $R$ approaches quickly 1 and scale invariances become increasingly good with 
increasing dimension~\cite{CSD16}, which is to be expected on the basis of this work.

 One may interpret the fact that in finite dimensions  the scaling properties hold to a good approximation~\cite{gnandyreIV,gnandyreV,BSD14}
  as a symptom of the high-dimensional approximation being qualitatively good. Note that this approximation
 is intimately tied to the Random First-Order Transition scenario for dense liquids~\cite{KW87,KT87,KT87b,KT89,KTW89,KT95,MP00,BB04}, so this is another instance in which we are confronted  with
 a  unifying perspective.\\

\paragraph*{\bf Acknowledgments --}

 We warmly thank Ludovic Berthier, Jeppe Dyre, Gilles Tarjus and Francesco Zamponi for insightful discussions. T. M. acknowledges funding from a CFM foundation grant.

 \clearpage
 
 \begin{widetext}
 
 \makeatletter
\let\toc@pre\relax
\let\toc@post\relax
\makeatother
 
\section*{Appendix}

Here we detail the derivation of the virial-energy correlation coefficient at first order in the virial expansion, equation~\eqref{eq:Rdio} of the main text. We use this expression to prove that $R=1$ in the case
of the exponential potential treated in this paper. We compute it as well in the special case of the Lennard-Jones potential at $O(1)$ temperature, which is used to get figure~\ref{fig:LJ}.

\subsection{Virial-energy correlations in high dimensions}\label{sub:R}

The free energy of the liquid and the pressure are respectively given by~\cite{WRF87,FP99,hansen,MH61}
\begin{equation}\label{eq:FP}
\begin{split}
  \frac{\b F}{N}&=\ln \r -1 -\frac{\r}{2}\int \dd \mathbf{r}\, \left(e^{-\b V(r)}-1\right)\\
   \frac{\b P}{\r}&=1 -\frac{\r}{2}\int \dd \mathbf{r}\, \left(e^{-\b V(r)}-1\right)
\end{split}
\end{equation}
where $\r=N/V$.\\
For hard spheres of diameter $\s$, the reduced pressure (for example) reads $\b P/\r=1 +\r \VV_d(\s)/2$ with $\VV_d(\s)=\Omega_d\s^d/d$ the $d$-dimensional volume of the sphere of diameter $\s$. \\
As mentioned in \cite[Appendix A]{gnandyreIV}, the existence of approximate isomorphs in finite dimension is equivalent to strong virial-energy correlations, measured by $R$, 
the Pearson correlation coefficient~\cite{gnandyreI,gnandyreII,BSD14},
\begin{equation}\label{eq:defR}
 R=\frac{\la \D W\D U\ra}{\sqrt{\la (\D W)^2\ra\la (\D U)^2\ra}}
\end{equation}
where $W$ is the microscopic virial defined as~\cite{gnandyreI,gnandyreII}:
\beq
W=-\frac 1d \sum_{i=1}^N \mathbf{r}_i\cdot\mathbf{\nabla}_{\mathbf{r}_i}U
\eeq
% This has been checked in simulations for a wide range of potentials~\cite{gnandyreI,gnandyreII,gnandyreIII,gnandyreIV,gnandyreV,BSD14,BD14}. For inverse-power law (IPL) potentials $V_{\rm IPL}(r)\propto1/r^n$, since $W\propto U$ are perfectly correlated, $R=1$ and isomorphs are exact even in finite dimension.\\
One can easily show~\cite[Appendix B]{gnandyreI} that equilibrium fluctuations can be computed 
from derivatives of equilibrium averages: 
\begin{equation}\label{eq:laUra}
  \frac{\partial\la W \ra}{\partial \b}=-\la WU\ra+\la W \ra \la U \ra=-\la \D W\D U\ra\,,\hskip15pt\la (\D U)^2\ra=-\frac{\partial^2 (\b F)}{ \partial \b^2}
\end{equation}
One can prove generically~\cite{hansen}, separating contributions from internal forces and forces on the walls, and using ergodicity, that $\b P/\r=1+\b\la W\ra/N$ (virial equation). Hence
from~\eqref{eq:FP} we can get 
\begin{equation}\label{eq:laWra}
\begin{split}
\la W\ra=&-\frac{N\r}{2\b}\int \dd \mathbf{r}\, \left(e^{-\b V(r)}-1\right)\\
 \frac{\partial\la W \ra}{\partial \b}=&-\frac{\r N}{2\b^2}\int \dd \mathbf{r}\, \left(e^{-\b V(r)}-1\right)-\frac{\r N}{2\b}\int \dd \mathbf{r}\, V(r)e^{-\b V(r)}
\end{split}
\end{equation}
$\la (\D U)^2\ra$ is computed using the expression of the liquid free energy~\eqref{eq:FP}. We now need an expression for $\la (\D W)^2\ra=\la W^2\ra-\la W \ra^2 $. Let us define the pair distribution function
\begin{equation}
 \r^{(2)}(\mathbf{x},\mathbf{y})=\sum_{i\neq j}\d(\mathbf{x}-\mathbf{r}_i)\d(\mathbf{y}-\mathbf{r}_j)
\end{equation}
Note that 
\begin{equation}
 \begin{split}
  \sum_i\mathbf{r}_i\cdot\mathbf{\nabla}_{\mathbf{r}_i}U&=\sum_{i\neq j}\mathbf{r}_i\cdot \mathbf{\nabla}_iV(|\mathbf{r}_i-\mathbf{r}_j|)=\frac12\sum_{i\neq j}\mathbf{r}_i\cdot \mathbf{\nabla}_iV(|\mathbf{r}_i-\mathbf{r}_j|)
  +\frac12\sum_{i\neq j}\mathbf{r}_j\cdot \mathbf{\nabla}_jV(|\mathbf{r}_j-\mathbf{r}_i|)\\
  &=\frac12\sum_{i\neq j}|\mathbf{r}_i-\mathbf{r}_j|V'(|\mathbf{r}_i-\mathbf{r}_j|)
 \end{split}
\end{equation}
therefore
\begin{equation}
 \la W\ra =-\frac{1}{2d}\int \dd\mathbf{x}\dd\mathbf{y}\, \la \r^{(2)}(\mathbf{x},\mathbf{y})\ra|\mathbf{x}-\mathbf{y}|V'(|\mathbf{x}-\mathbf{y}|)=-\frac{N}{2d}\r\int\dd\mathbf{r}\, rg(r)V'(r)
\end{equation}
where $g(r)$ is the radial distribution function~\cite{hansen}. The lowest order virial contribution to the radial distribution function is $g(r)=e^{-\b V(r)}$~\cite{hansen,MH61}. This is consistent with~\eqref{eq:laWra}
since with an integration by parts
\begin{equation}
\begin{split} 
\frac1d\int\dd\mathbf{r}\, rg(r)V'(r)=&\frac{\Omega_d}{d}\int\dd r\, r^de^{-\b V(r)}V'(r)=-\frac{\Omega_d}{\b d}\underbrace{\left[(e^{-\b V(r)}-1)r^d\right]_0^\io}_{=0}
+\frac{\Omega_d}{\b}\int\dd r\, r^{d-1}\left(e^{-\b V(r)}-1\right)\\
=&\frac1\beta\int\dd\mathbf{r}\,\left(e^{-\b V(r)}-1\right)
\end{split}
\end{equation}
Similarly,
\begin{equation}
\begin{split}
 \la W^2\ra&=\frac{1}{4d^2}\int\dd\mathbf{x}_1\dd\mathbf{x}_2\dd\mathbf{x}_3\dd\mathbf{x}_4\, \la \r^{(2)}(\mathbf{x}_1,\mathbf{x}_2)\r^{(2)}(\mathbf{x}_3,\mathbf{x}_4)\ra|\mathbf{x}_1-\mathbf{x}_2||\mathbf{x}_3-\mathbf{x}_4|V'(|\mathbf{x}_1-\mathbf{x}_2|)V'(|\mathbf{x}_3-\mathbf{x}_4|)\\
 &=\frac{V^2}{4d^2}\int\dd\mathbf{x}\dd\mathbf{y}\la \r^{(2)}(\mathbf{0},\mathbf{x})\r^{(2)}(\mathbf{0},\mathbf{y})\ra|\mathbf{x}||\mathbf{y}|V'(|\mathbf{x}|)V'(|\mathbf{y}|)
 \end{split}
\end{equation}
To get $\la \r^{(2)}(\mathbf{0},\mathbf{x})\r^{(2)}(\mathbf{0},\mathbf{y})\ra$ we will use the smallest order in its virial expansion. 
Working with the grand-canonical partition function, we have~\cite{MH61}
\begin{equation}\label{eq:Xi}
 \begin{split}
  \Xi&=\sum_{N\geqslant0}\frac{e^{\b\m N}}{N!}\int \dd\mathbf{r}_1\dots\dd\mathbf{r}_N\prod_{i<j}e^{-\b V(\mathbf{r}_i,\mathbf{r}_j)}
  =\sum_{N\geqslant0}\frac{e^{\b\m N}}{N!}\int \dd\mathbf{r}_1\dots\dd\mathbf{r}_N e^{-\frac\b2 \int \dd\mathbf{x}\dd\mathbf{y}\,\r^{(2)}(\mathbf{x}, \mathbf{y}) V(\mathbf{x},\mathbf{y})  }\\
\ln\Xi &=N(1-\ln\r-\b\m)+\frac12\int\dd\mathbf{x}\dd\mathbf{y}\,\r(\mathbf{x})\r(\mathbf{y})\left(e^{-\b V(\mathbf{x},\mathbf{y})}-1\right)%\begin{tikzpicture}[baseline={([yshift=-.5ex]current bounding box.center)}] \draw[line width=1pt] (0,0) -- (1,0); \draw[fill=black] (0,0) circle (0.1) ;\draw[fill=black] (1,0) circle (0.1) ; \end{tikzpicture} + \begin{tikzpicture}[baseline={([yshift=-.5ex]current bounding box.center)}] \draw[line width=1pt] (0:0) -- (60:1) -- (0:1) -- (0:0);\draw[fill=black] (0:0) circle (0.1) ;\draw[fill=black] (60:1) circle (0.1) ;\draw[fill=black] (0:1) circle (0.1) ;\end{tikzpicture} + \begin{tikzpicture}[baseline={([yshift=-.5ex]current bounding box.center)}] \draw[line width=1pt] (0,0) -- (0,1) -- (1,1) -- (1,0) -- (0,0);\draw[fill=black] (0,0) circle (0.1) ;\draw[fill=black] (1,0) circle (0.1) ;\draw[fill=black] (0,1) circle (0.1) ;\draw[fill=black] (1,1) circle (0.1) ;\end{tikzpicture} + \begin{tikzpicture}[baseline={([yshift=-.5ex]current bounding box.center)}] \draw[line width=1pt] (0,0) -- (0,1) -- (1,1) -- (1,0) -- (0,0) -- (1,1) ;\draw[fill=black] (0,0) circle (0.1) ;\draw[fill=black] (1,0) circle (0.1) ;\draw[fill=black] (0,1) circle (0.1) ;\draw[fill=black] (1,1) circle (0.1) ;\end{tikzpicture} + \begin{tikzpicture}[baseline={([yshift=-.5ex]current bounding box.center)}] \draw[line width=1pt] (0,0) -- (0,1) -- (1,1) -- (1,0) -- (0,0) -- (1,1) ; \draw[line width=1pt] (1,0) -- (0,1) ; \draw[fill=black] (0,0) circle (0.1) ;\draw[fill=black] (1,0) circle (0.1) ;\draw[fill=black] (0,1) circle (0.1) ;\draw[fill=black] (1,1) circle (0.1) ;\end{tikzpicture} +  \dots 
 \end{split}
\end{equation}
%with $\r$ nodes and $f(\mathbf{x},\mathbf{y})=e^{-\b V(|\mathbf{x}-\mathbf{y}|)}-1$ links (Mayer function). 
where here $\r(\mathbf{x})=\la\sum_i\d(\mathbf{x}-\mathbf{r}_i)\ra$ is the equilibrium local density as in liquid theory~\cite{hansen}. Each line of~\eqref{eq:Xi} is respectively used to derive the corresponding line below:
\begin{equation}\label{eq:dxi}
\begin{split}
 \frac{\d^2 \ln \Xi}{\d V(\mathbf{x},\mathbf{y})\d V(\mathbf{r},\mathbf{r'})}=&\frac{\b^2}{4}\left[\la \r^{(2)}(\mathbf{x},\mathbf{y})\r^{(2)}(\mathbf{r},\mathbf{r'})\ra-\la \r^{(2)}(\mathbf{x},\mathbf{y})\ra\la\r^{(2)}(\mathbf{r},\mathbf{r'})\ra\right]\\
 &=\frac{\b^2}{2}\r(\mathbf{x})\r(\mathbf{y})\d(\mathbf{x}-\mathbf{r})\d(\mathbf{y}-\mathbf{r'})e^{-\b V(\mathbf{x},\mathbf{y})}
\end{split}
\end{equation} 
Multiplying~\eqref{eq:dxi} by $|\mathbf{x}-\mathbf{y}||\mathbf{r}-\mathbf{r'}|V'(|\mathbf{x}-\mathbf{y}|)V'(|\mathbf{r}-\mathbf{r'}|)$ and integrating over $\mathbf{x}$, $\mathbf{y}$, $\mathbf{r}$ and 
$\mathbf{r'}$ we can make the connection with $\la(\D W)^2\ra$ and obtain\footnote{Here we considered only the first term of the virial expansion, but one may wonder if doing the second derivatives with respect to 
the potential used here and then using the infinite-dimensional limit affects this truncation, leading to consider higher-order diagrams. One can check, for example with the triangle term 
\begin{tikzpicture}[baseline={([yshift=-.5ex]current bounding box.center)}] \draw[line width=1pt] (0:0) -- (60:0.35) -- (0:0.35) -- (0:0);
\draw[fill=black] (0:0) circle (0.075) ;\draw[fill=black] (60:0.35) circle (0.075)  ;\draw[fill=black] (0:0.35) circle (0.075)  ;\end{tikzpicture},
that it does not since the factors $|\mathbf{x}-\mathbf{y}||\mathbf{r}-\mathbf{r'}|V'(|\mathbf{x}-\mathbf{y}|)V'(|\mathbf{r}-\mathbf{r'}|)$ 
play a similar role to reintroducing the missing Mayer functions due to derivation.}
\begin{equation}\label{eq:DW}
\la(\D W)^2\ra= \la W^2\ra - \la W\ra^2=\frac{2\r^2}{4d^2}\int\dd\mathbf{x}\dd\mathbf{y}\,e^{-\b V(|\mathbf{x}-\mathbf{y}|)}(\mathbf{x}-\mathbf{y})^2V'(|\mathbf{x}-\mathbf{y}|)^2
=\frac{N\r}{2d^2}\int\dd\mathbf{r}\,r^2V'(r)^2e^{-\b V(r)}
\end{equation} 
All in all,~\eqref{eq:FP},~\eqref{eq:defR},~\eqref{eq:laUra} and~\eqref{eq:DW} gives for the correlation coefficient
\begin{equation}\label{eq:Rio}
 R\underset{d\to\io}{\sim}-d\frac{\int\dd\mathbf{r}\,\left(e^{-\b V(r)}-1\right)+\b\int\dd\mathbf{r}\,V(r)e^{-\b V(r)}}{\b^2\sqrt{\int\dd\mathbf{r}\,r^2V'(r)^2e^{-\b V(r)}\int\dd\mathbf{r'}\,V(r')^2e^{-\b V(r')}}}
=-\frac{\int\dd\mathbf{r}\,[rV'(r)+dV(r)]e^{-\b V(r)}}{\b\sqrt{\int\dd\mathbf{r}\,r^2V'(r)^2e^{-\b V(r)}\int\dd\mathbf{r'}\,V(r')^2e^{-\b V(r')}}}
\end{equation}
through an integration by parts. This is~\eqref{eq:Rdio} of the main text. 

\subsection{Case of the exponential potential}

We can write it with explicit inverse temperature factors:
\begin{equation}
 R=-\frac{\int\dd\mathbf{r}\,[r\beta V'(r)+d\beta V(r)]e^{-\b V(r)}}{\sqrt{\int\dd\mathbf{r}\,r^2\beta^2 V'(r)^2e^{-\b V(r)}\int\dd\mathbf{r'}\,\b^2 V(r')^2e^{-\b V(r')}}}
\end{equation}

% Note that 
% \begin{equation}
%  \begin{split}
%   \int\dd\mathbf{r}\,V(r)e^{-\b V(r)}&=-\frac{\partial}{\partial\b}\int\dd\mathbf{r}\,\left(e^{-\b V(r)}-1\right)\\
%   \int\dd\mathbf{r}\,V(r)^2e^{-\b V(r)}&=\frac{\partial^2}{\partial\b^2}\int\dd\mathbf{r}\,\left(e^{-\b V(r)}-1\right)
%  \end{split}
% \end{equation}
% Let us check if $R=1$ for IPL potentials where this statement is exact $\forall d$. Through an integration by parts
% \begin{equation}\label{eq:Rdio}
%  R=-\frac{\int\dd\mathbf{r}\,(rV'(r)+dV(r))e^{-\b V(r)}}{\b\sqrt{\int\dd\mathbf{r}\,r^2V'(r)^2e^{-\b V(r)}\int\dd\mathbf{r'}\,V(r')^2e^{-\b V(r')}}}
% \end{equation}
% For IPL potentials, $V_{\rm IPL}=\epsilon/(r/\s)^{da}$ and $rV_{\rm IPL}'(r)=-da V_{\rm IPL}(r)$ so
% \begin{equation}
%  R_{\rm IPL}=\frac{a-1}{a}\frac{\int\dd\mathbf{r}\, V(r)e^{-\b V(r)}}{\b\int\dd\mathbf{r}\, V(r)^2e^{-\b V(r)}}=-\frac{a-1}{a}\frac{\G(-1/a)}{a\G(2-1/a)}=1
% \end{equation}
% with a change of variables to cast in the form of the Gamma function $\G(x)=\int_0^\io\dd t\, t^{x-1}e^{-t}$ and using $\G(x+1)=x\G(x)$.\\

Here the situation is similar to the Mayer integral, with the potential $\b V=\b e^{-d A}$ playing the role of the Mayer function $e^{-\b V}-1$. We can make a similar analysis as in the main text, comparing contributions
of an integrand of the type $r^d\b V(r)e^{-\b V(r)}$:
\begin{equation}
\begin{split}
 \bullet ~\textrm{for~}r<\rs:&\hskip15pt r^d\b V(r)e^{-\b V(r)}=r^d\wt\b e^{d(A(\rs)-A(r))} \exp[-\wt\b e^{d(A(\rs)-A(r))}]\\
 \bullet ~\textrm{for~}r\sim \rs:&\hskip15pt (\rs)^d\wt\b e^{-\wt\b}\\
 \bullet ~\textrm{for~}r>\rs:&\hskip15pt r^d\b V(r)e^{-\b V(r)}\sim r^d\wt\b e^{d(A(\rs)-A(r))}
\end{split}
\end{equation}
Still under the same hypothesis that $\left[\ln r-A(r)\right]_r^{\rs}>0$, the first regime $r<\rs$ is strongly damped by the Boltzmann factor, and the other two regimes $r\sim\rs$ and $r>\rs$ compares exactly as 
for the Mayer integral in the main text, where the same conditions apply. The other term in the numerator of~\eqref{eq:Rio} involves $rV'(r)$ and is treated the same way. In the denominator, the analysis
is the same except that due to the power 2 we will need the condition that $\ln r-2A(r)$ decreases instead, for the region around $\rs$ to dominate the integral; since this means that $A'(r)>1/2r$ it is less
constraining than the previous condition $\ln r-A(r)$ decreases\footnote{Actually the condition is  $\left[\ln r-A(r)\right]_r^{\rs}>0$ which is weaker than 
$\ln r-A(r)$ decreases, but this is to fix ideas.}, so it is automatically fulfilled with the latter condition. Therefore we expand the integral involved in~\eqref{eq:Rio} as we did 
in the main text: we set $r=\rs(1+\rt/d)$, and around $r=\rs$ we have
\begin{equation}
 r \b V'(r) =-drA'(r)\b V(r)\sim -d\rs A'(\rs)\b V(\rs) e^{-\rt/\a}=-d\frac{\wt\b}{\a}e^{-\rt/\a}\hskip15pt\textrm{and}\hskip15pt\b V(r)\sim \wt\b e^{-\rt/\a}
\end{equation}
so that the virial-energy correlation coefficient becomes
\begin{equation}
\begin{split}
 R&\underset{d\to\io}{\sim}\frac{1-\a}{\wt\b}\frac{\int_{-\io}^\io\dd\rt\, e^{\rt(1-1/\a)-\wt\b\exp(-\rt/\a)}}{\int_{-\io}^\io\dd\rt\, e^{\rt(1-2/\a)-\wt\b\exp(-\rt/\a)}}\\
 &\underset{x\equiv\wt\b e^{-\rt/\a}}{=}(1-\a)\frac{\int_0^\io\dd x\,x^{-\a} e^{-x}}{\int_0^\io\dd x\,x^{1-\a} e^{-x}}=(1-\a)\frac{\G(1-\a)}{\G(2-\a)}=1
\end{split}
\end{equation}
We conclude that all liquids within the class of potentials considered here are strongly correlated in high dimension, which is expected in light of~\cite{gnandyreIV} since we find the existence of 
exact isomorphs.\\

\subsection{Case of the Lennard-Jones potential}

The Lennard-Jones potential in $d$ dimensions reads:
\begin{equation}\label{eq:LJ}
 V_{\rm LJ}(r)=\epsilon\left[\left(\frac{r}{\sigma}\right)^{-4d}-\left(\frac{r}{\sigma}\right)^{-2d}\right]
\end{equation}
In this case the Mayer function is very similar to the previous case: it is $-1$ at short distance, then there is an $O(1)$ part over a range $O(1/d)$ around $\s$ where there is a positive bump, and it is exponentially
small at large distances (see~\figref{fig:Mayer}).

\begin{figure}[h!]
\begin{tabular}{c}
 \includegraphics[width=10cm]{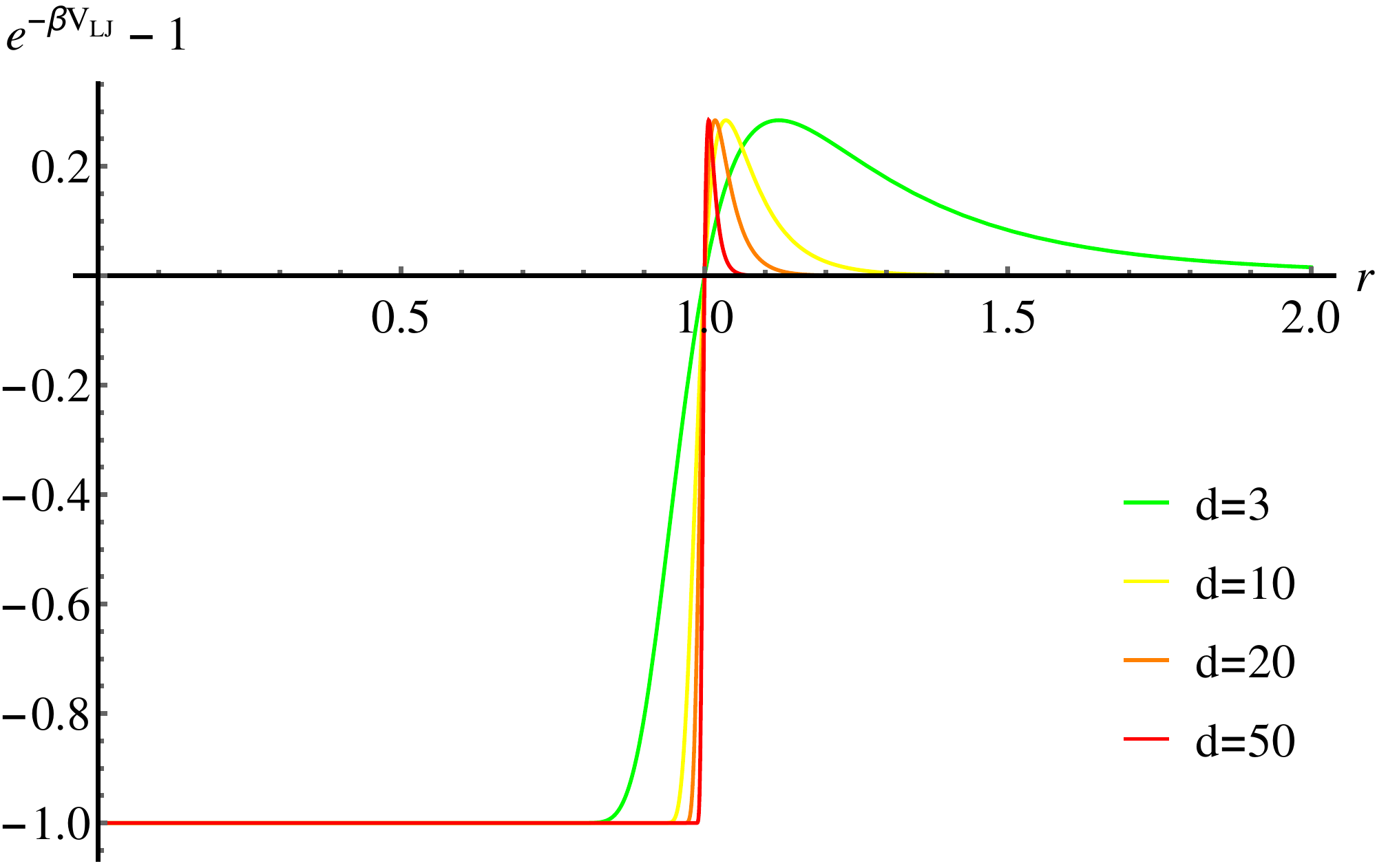} \\(a)
  \includegraphics[width=10cm]{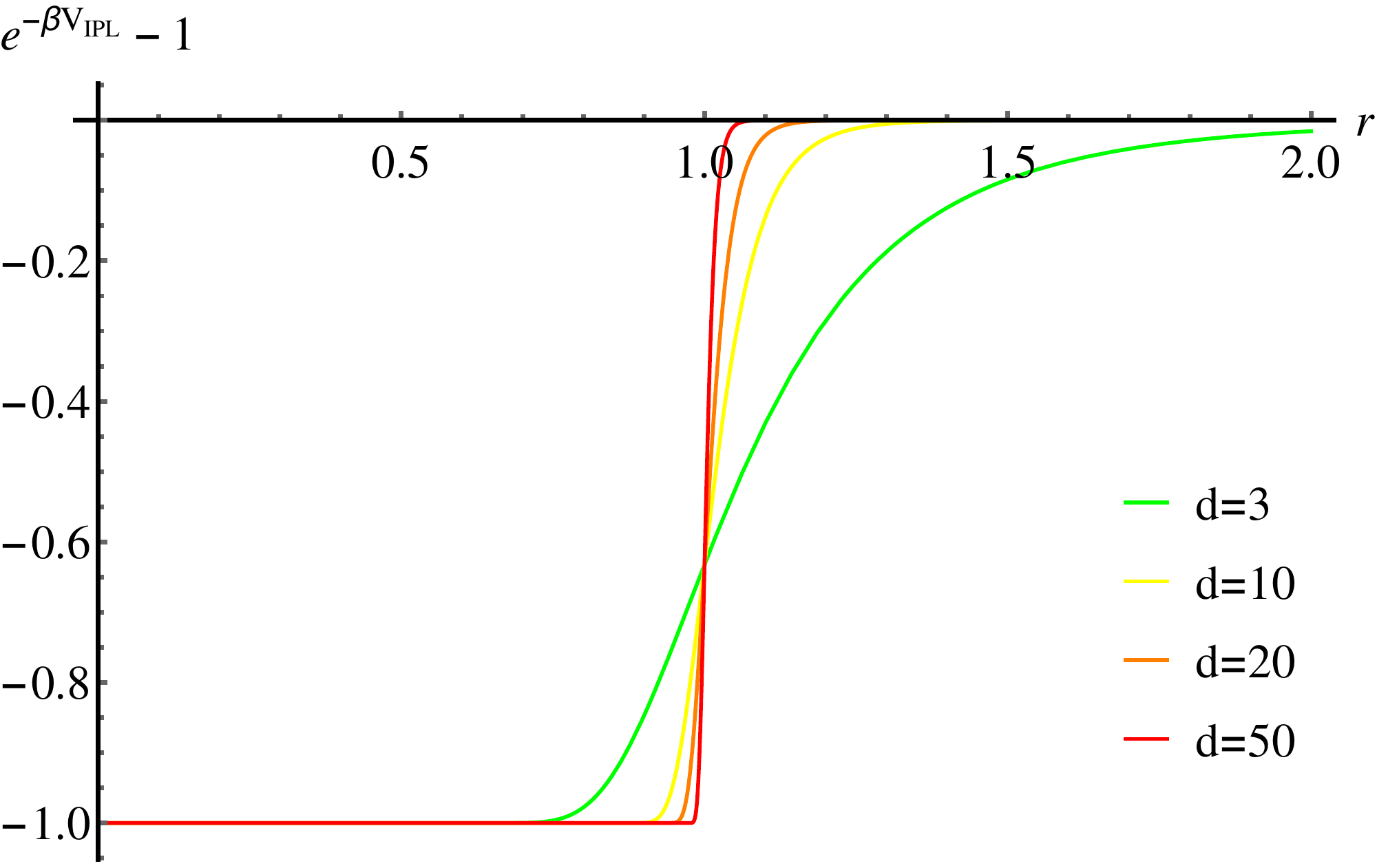}\\ (b)
\end{tabular}
 \caption{(a) The Lennard-Jones Mayer function for $\b\epsilon=1$, $\s=1$ and $d=3,10,20,50$. (b) For comparison, the Mayer function for an IPL potential, similar to figure~\ref{fig:vir},  
 $V_{\rm IPL}=\epsilon (\s/r)^{d/\a}$ with $\b\epsilon=1$, $\s=1$, $\a=1/2$, and $d=3,10,20,50$. 
As $d$ increases one clearly sees the emergence of the three distinct regions of the Mayer function: exponentially close to $-1$ for $r<\s$, $O(1)$ for $r$ around $\s$ with 
$O(1/d)$ fluctuations, and exponentially small for $r>\s$.}
 \label{fig:Mayer}
\end{figure}

We focus on the peculiar regime is where the temperature is $O(1)$. Then $r^*$ is defined in the $O(1/d)$ region close to $\sigma$ where attractive and repulsive parts compete; 
its precise value does not matter\footnote{Since $V(\sigma)=0$, only at this point, it is wise not to use it to define $\rs$ but any other in this $O(1/d)$ range, such as the minimum.}. 
We can expand the potential once again setting $r=\rs(1+\rt/d)$, giving an effective potential
\begin{equation}
 \beta V_{\rm LJ}(r)\sim\widetilde\beta V_{\rm LJ}^{\rm eff}(\rt)=\widetilde\beta(e^{-4\rt}-e^{-2\rt})
\end{equation}
We now compute the virial-energy correlation coefficient in this regime, as in~\secref{sub:R} by computing the dominant contribution (around $\rs$) of the 
formula~\eqref{eq:Rio}, giving
\begin{equation}
 R_{\rm LJ}\underset{d\to\io}{\sim}-\frac{1}{\widetilde\beta}\frac{\int_{-\io}^\io\dd\rt\,(-3e^{-4\rt}+e^{-2\rt}) e^{\rt-\wt\b(e^{-4\rt}-e^{-2\rt})}}
 {\sqrt{\int_{-\io}^\io\dd\rt\,(-4e^{-4\rt}+2e^{-2\rt})^2 e^{\rt-\wt\b(e^{-4\rt}-e^{-2\rt})}\int_{-\io}^\io\dd\rt\,(e^{-4\rt}-e^{-2\rt})^2 e^{\rt-\wt\b(e^{-4\rt}-e^{-2\rt})}}}
\end{equation}
This formula is used to plot $R_{\rm LJ}$ as a function of $\wt\b$ in figure~\ref{fig:LJ}.
  
\clearpage 

 \end{widetext}

\bibliographystyle{mioaps}
\bibliography{HS}

\end{document}